\date{}
\documentclass[showpacs,twocolumn]{revtex4}
\usepackage{graphicx,psfrag,amsmath,amssymb,amsfonts,bbm,latexsym,color,dcolumn,epsf}

\begin{document}

\title{Decoherence induced by a composite environment}

\author{Fernando C. Lombardo \footnote{Electronic address: lombardo@df.uba.ar}}
\author{Paula I. Villar \footnote{Electronic address: paula@df.uba.ar}}
\affiliation{Departamento de F\'\i sica, FCEyN UBA,
Facultad de Ciencias Exactas y Naturales, Ciudad Universitaria,
Pabell\' on I, 1428 Buenos Aires, Argentina}

\date{today}

\begin{abstract}
We study the decoherence induced by the environment over a composite
quantum system, comprising two coupled subsystems A and B, which
may be a harmonic or an upside-down oscillators. We analyze the case in
which the B-subsystem is in direct interaction with a thermal
bath, while the other remains isolated from the huge reservoir. We
compare the results concerning the decoherence suffered by the
A-subsystem.  
\end{abstract}


\maketitle

\newcommand{\beq}{\begin{equation}}
\newcommand{\eeq}{\end{equation}}
\newcommand{\beqa}{\begin{eqnarray}}
\newcommand{\eeqa}{\end{eqnarray}}
\newcommand{\beqas}{\begin{eqnarray*}}
\newcommand{\eeqas}{\end{eqnarray*}}

Decoherence is the process by which most pure states evolve into
mixtures due to the interaction with an environment
\cite{zurek-gral}. The very notion of a quantum open system implies
the appearance of dissipation and decoherence as an ubiquitous
phenomena and plays important roles in different branches of physics
\cite{more,JPP-LH}. Oftentimes, a large system, consisting of two or a few subsystems
interacting with their environment (thermal
bath comprising a large number of degrees of freedom), can be
adequately described as a composite system. Examples include
electron transfer in solution \cite{prezhdo}, a large biological
molecule, vibrational relaxation of molecules in solution, excitons
in semiconductors coupled to acoustic or optical phonon modes.
Quantum processes in condensed phases are usually studied by
focusing on a small subset of degrees of freedom and considering the
other degrees of freedom as a bath.

In this article, we analyze the decoherence induced in a composite
quantum system, in which an observer can distinguish between two
different subsystems, one of them coupled to an external
environment. Our composite system is formed by a subsystem A coupled
to a subsystem B which is also bilinearly coupled to an external environment $\cal{E}$.
The coupling to this external environment is only through subsystem B.
Subsystem A remains isolated from $\cal{E}$ but for the information delivered
by B through a bilinearly coupling between subsystems A and B. We will consider
the thermal bath to be at high temperature and will work in the underdamped limit.

In order to investigate this problem we mainly consider a simple
model where subsystem A is represented by a harmonic oscillator and
subsystem B is an upside-down one. The main motivation for studying
this model is to deepen and enlighten previous analysis of decoherence induced by chaotic
environments. The upside-down oscillator has  recently been used to
model a chaotic environment which induces decoherence on the system
\cite{robin}. Even though it is an oversimplified model for a
chaotic environment, it displays exponential sensitivity to
perturbations, which is crucial in order to analyze chaotic
evolutions. In this context, we shall
consider two different cases. Firstly, the case where the chaotic degree of
freedom is part of the environment (i.e. an unstable system B) and
is directly coupled to an external reservoir $\cal{E}$ and to another
subsystem A with different bare frequency. Secondly, the case where
subsystem A is
unstable and directly coupled to a harmonic oscillator (subsystem B)
which, in turn, is coupled to an external bath $\cal{E}$. These are the
extension of previous works done in \cite{robin} and \cite{diana,elze}
for the first and second case, respectively. In both situations, we
will estimate the decoherence time, which is the usual scale after
which classicality emerges. 

The analysis is completed by the inclusion of the other two
different possibilities for the quantum composite system, i.e., a
composite system constituted by a subsystem A coupled to subsystem
B, both harmonic oscillators, and a composite system formed by
subsystem A coupled to subsystem B, both inverted oscillators. As in
the other two cases mentioned above, subsystem B is also coupled to
an external reservoir $\cal{E}$. All in all, we have four different
composite systems to analyze. 
For every and each situation, we study the dynamics of the subsystem
A. Not only did we study the influence of ``its" environment (formed
by subsystem B and $\cal{E}$) at low and high temperature but also
in the absence of the external reservoir $\cal{E}$.  Each case
develops a different dynamics, being possible, in some cases, to
find a quantum open system described using mixed quantum-classical
dynamics \cite{kapral,kapral1, anu} (part of the composite system completely
decohered and the other does not).

The total AB${\cal E}$ classical action is 
$S[x,q,Q] = S_{\rm A}[x] + S_{\rm B}[q] + S_{\rm {\cal E}}[Q] +
S_{\rm AB}[x,q] + S_{B \cal E}[q,Q]$.
In the spirit of the quantum Brownian motion (QBM) paradigm, the environment is
taken to be a set of $N$ independent harmonic oscillators with
frequencies $\tilde{\omega }_n$, masses $m_n$, and coordinates $\hat{Q}_n$. 
Subsystem B consists of a single oscillator (upside-down or
harmonic, depending on the case considered) with bare mass
$M_{\rm B}$, frequency $\Omega$ and coordinate operator $\hat{q}$.
The interaction between subsystem B and the thermal environment is assumed to be
bilinear $q(s) Q_n(s)$. For simplicity, we assume an Ohmic environment,
with the spectral density $I_{{\cal E}}(\omega ) = 2 M \gamma_0 \tilde{\omega}
e^{{-{\tilde{\omega}^2\over{\Lambda^2}}}}$, where $\Lambda$ is a
physical cutoff, related to the maximum frequency present in the environment. 
Finally, we consider subsystem A consisting of a single
oscillator (again, this oscillator can be an upside-down or harmonic one)
with coordinate operator $\hat{x}$ and frequency $\omega$. 
We suppose that subsystem A is bilinearly coupled to subsystem B
by the interaction term $\lambda x(s) q(s)$. 

The dynamical properties of interest can be computed from the
density matrix of the system at time $t$. However, if we
want to know how is the decoherence process for the subsystem A,
we have to trace over all the degrees of freedom that belong to a 
composite environment. We can assume that our new problem is a subsystem A
and a subsystem B which are coupled through an effective
interaction defined by
\beqa &&S_{\rm eff}^{\rm int}(x,q,x',q') = S_{\rm AB}(x,q)- S_{\rm AB}(x',q')
\nonumber \\ 
&&- 2 M_{\rm B} \gamma_0 \int_0^t ds \Delta q(s) \dot{\Sigma} q(s)\nonumber \\ 
&&+ i\frac{2M_{\rm B}\gamma_0k_B T}{\hbar} \int_0^t ds
(\Delta q(s))^2,\eeqa 
where the last two terms are the 
usual influence action for the QBM problem, in the environmental 
high-T limit \cite{fey,Hu,Grabert,habib}.

After integrating out the external bath, the information about 
subsystem A is encoded in the reduced density matrix. That is to say, we have yet  
to trace over the degrees of freedom of subsystem B. This final reduced density matrix 
satisfy a master equation which can be presented, as usual, as \cite{Hu,hu2}
\beqa &i \hbar \dot{\rho}_{\rm r}& (x,x';t)=\bigg[-\frac{\hbar^2}{2 M_{\rm A}} \bigg[
\frac{\partial^2}{\partial x^2}-\frac{\partial^2}{\partial x'^2}\bigg]  \nonumber \\ &+ &
\frac{1}{2} M_{\rm A}\Omega^2(x^2-x'^2)\bigg] \rho_{\rm r}(x,x';t) \nonumber \\ &+ &
\frac{1}{2}M_{\rm A} \delta\Omega^2(t)
(x^2-x'^2) \rho_{\rm r}(x,x';t) \nonumber \\ &- &
 i \hbar \Gamma(t)(x-x')\bigg[\frac{\partial}
{\partial x}-\frac{\partial}{\partial x'}\bigg]  \rho_{\rm r}
(x,x';t) \nonumber \\ &- & i M_{\rm A} {\cal D}(t)
(x-x')^2 \rho_{\rm r}(x,x';t) \nonumber \\ &- &
\hbar \Gamma(t) \textit{f}(t)(x-x')\bigg[\frac{\partial}
{\partial x}+\frac{\partial}{\partial x'}\bigg] \rho_{\rm r}(x,x';t)
\label{master},\eeqa
where ${\cal D}(t)(x-x')^2$ is the new diffusion term, which produces the decay of
the off-diagonal elements. For simplicity we omitted the subindex ${\rm f}$ to
indicate the final configuration $x_{\rm f}$.
The total diffusion coefficient is given by

\beqa \lefteqn{{\cal D}(t) =\frac{2 \gamma_0 k_B T}{\hbar \Omega^2}
\lambda^2
\int_0^t ds ~\Delta q_{\rm cl}(s)~ \dot{\Delta} q_{\rm cl}(s) }
~~~~~~~~~~~~~~~\nonumber \\
&+ &  \frac{\lambda^2\sigma}{32 \hbar} \int_0^t ds ~\tilde{\nu}(t-s)
~\Delta x_{\rm cl}(s) \label{D},\eeqa where $\Delta q_{\rm cl}(s) = q_{\rm cl} - 
{{q}\prime}_{\rm cl}$ is built from the solution of 
$\ddot q(s) - \Omega^2 q(s)=\frac{\lambda}{M_{\rm B}} x(s)$ (assuming 
subsystem B is an upside-down oscillator (case (a))). 
After imposing
initial and final conditions $q(s=0)=q_0$ and $q(s=t)=q_{\rm f}$,
respectively, we write the complete classical solution as \beqa
\lefteqn{q_{\rm cl}(s)=q_0\frac{\rm {\sinh}(\Omega(t-s))}{\rm
{\sinh}(\Omega t)}+ q_{\rm f} \frac{\rm {\sinh}(\Omega s)}{\rm
{\sinh}(\Omega t)}} \nonumber \\ & &-\frac{\lambda}{M_{\rm B} \Omega}
\frac{\rm {\sinh}(\Omega s)}{\rm {\sinh}(\Omega t)} \int_0^t x(u)
\rm{\sinh}(\Omega (s-u)) du \nonumber
\\ & & + \frac{\lambda}{M_{\rm B} \Omega}\int_0^s x(u) \rm{\sinh}(\Omega
(s-u)) du. \label{classol}\eeqa
The kernel $\tilde{\nu}$ is the
new noise kernel (product of the interaction between subsystem A and B), and it is given by
$\tilde{\nu}(s_1-s_2)=\frac{\lambda^2\sigma}{32 \hbar}
\rm{\cosh}(\Omega(s_1-s_2))$. $\sigma$ is the width of the initial wave 
packet, used to describe the initial state of subsystem B. It is important to note that
$\Delta q_{\rm cl}(s)$ is the solution of the coupled system, and
the new noise kernel is not the usual T-dependent noise kernel of
the QBM problem. 

At this stage, we assume that our
subsystem A is a harmonic oscillator, (being possible to obtain
the solution for an upside-down oscillator by just replacing
$\omega$ for $i \omega$). If we ask for initial and final conditions of the form
$x(s=0)=x_0$ and $x(s=t)=x_{\rm f}$, the classical solution of the 
free equation is: $x_{\rm cl}(s)=x_0 \frac{\rm {\sin}(\omega(t-s))}{\rm {\sin}(\omega t)} +
x_{\rm f} \frac{\rm{\sin}(\omega s)}{\rm {\sin}(\omega t)}$.  

After integrating out all the degrees of freedom corresponding to
the external hot environment $Q_n$, and the coordinates $q$
belonging to the subsystem B, we obtained the diffusive terms that
induce decoherence on subsystem A. Therefore, we numerically
integrated the diffusive terms in time, in order to plot the
decoherence factor $\Gamma (t) = \exp\left\{- \int_0^t {\cal D}(s) ~ ds\right\}$.  
We will consider four different situations:
{\bf Case ({\mbox a}): Harmonic Oscillator + Upside-Down Oscillator + ${\cal E}$}. 
This is the generalization of the toy model
considered in Ref.\cite{robin} where they did not consider the
interaction of subsystem B (upside-down 
oscillator) with an external environment. It is easy to find results of \cite{robin}
just by setting $\gamma_0 = 0$ in our results. Case ({\mbox a})
is the situation in which a Brownian particle (in a harmonic
potential) suffers decoherence from an environment with one (or
more) chaotic degrees of freedom. {\bf Case ({\mbox b}): 
Upside-Down  Oscillator + Harmonic Oscillator + ${\cal E}$}. It represents the
possibility of studying decoherence induced on an unstable system
(toy model for a chaotic subsystem) by a completely harmonic
environment \cite{diana,pau,ZP,elze}. We will see that this is the most
decoherent system among all four cases studied in this
paper. {\bf Case ({\mbox c}): Harmonic Oscillator + Harmonic Oscillator + ${\cal E}$}. 
For completeness, we also consider the case of two harmonic
oscillators coupled together and one of them coupled to 
an external environment. {\bf Case ({\mbox d}): 
Upside-Down  Oscillator + Upside-Down Oscillator +
${\cal E}$}. We will see that this case is
the most sensitive to external perturbations (both subsystems are
unstable) when there is no external environment, thus decoherence is
much more effective than in the other cases. In particular, it is
interesting to note that this case decoheres long before the others
when there is no thermal environment ($\gamma_0=0$).

\begin{figure}[h]
\begin{center}
\includegraphics[width=8cm]{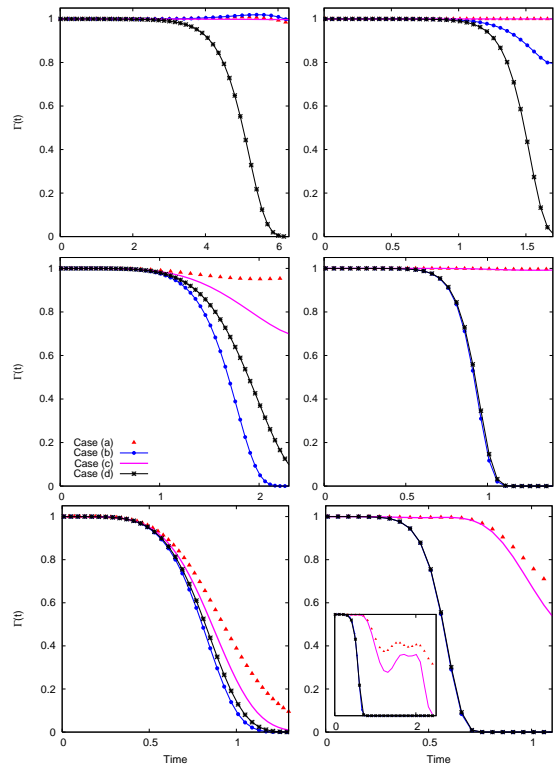}
\caption{Decoherence factor  $\Gamma (t)$.  
Isolated composite system decoheres first for the (d)
case (figures on the top). Plots in the middle and bottom are the decoherence factors for 
$\gamma_0 k_BT= 1$ and $100$, respectively. Case (b) is more decoherent. On the left, both 
frequencies of the subsystems A and B are of the same
order of magnitude ($\omega \approx \Omega$) and on the right we show the case 
$\omega > \Omega$} 
\label{figure}
\end{center}
\end{figure}

In order to illustrate the different behaviours, we present all four $\Gamma$ coefficients for two 
different situations: both frequencies of the subsystems A and B are of the same
order of magnitude ($\omega \approx \Omega$) and  when
 $\omega > \Omega$, as shown in Fig.\ref{figure}. 
Both cases are considered in the absence of external environment
$\cal{E}$ (i.e. $\gamma_0 = 0$) and for low and high values of
$\gamma_0 k_B T$.

From the numerical results shown on top of Fig.\ref{figure}, we can stress 
that in the absence of a hot bath, the
decoherence time is smaller in case (d) than in (b), and both of
them decohere long before cases (a) and (c). This is due to the fact
that subsystem A, which is the solely coupling to subsystem B,
generates noise and dissipation at large scales. Thus,
this noise and dissipation is bigger when the subsystem B is an
upside-down oscillator (case (d)) than when it is 
a harmonic oscillator (case (b)). In this situation ($\gamma_0=0$), 
case (d) is twofold exponential in time. In the 
former, the oscillatory dynamics of the A-oscillator and the 
hyperbolic stretching of the B-environment, proceed largely
independently of one another. The B-environment induces only minor
perturbations in the subsystem A and this subsystem  does not
disturb the environment. The stretching of the environment (due to
being an inverted oscillator) along its unstable manifold is
reflected in the system as diffusion. The same physical process
occurs in case (b), with the sole and essential difference that
the one stretching along an unstable direction is the subsystem A,
while the environment is oscillating. As this stretching results
in diffusion, the more stretching the system has, the more
diffusion it feels. Case (d) is the best example in this
``isolated'' model, because both, A and B, stretch along a
direction in the phase space, producing exponential diffusion.
This is the reason why it is the most decoherent case. Case (c) is
shown for completeness, but it is easily seen that decoherence
occurs in a longer time scale (there is no stretching here).
Therefore, case (d) decoheres at the time in which all the other
examples do not. 

As soon as the interaction between B and the thermal environment is
switched on, oscillator B dissipates not only on the bath but also
on A. This is shown in the middle and at the bottom of Fig.
\ref{figure}. At very high environmental
temperatures, there is no difference between cases (b) and (d); both
of them decohere in the same temporal scale. The huge reservoir dominates 
the diffusion coefficient (first term in Eq.(\ref{D}) . But they still differ from the cases
where there are harmonic oscillators as subsystems A (cases (a) and
(c)). The inset in Fig.\ref{figure} (at the right bottom), presents the behaviour of the
$\Gamma (t)$ factors for these cases for a longer time scale.
We can observe that we need to wait longer times for decoherence to
be effective in cases (a) or (c) with respect to (b) and (d) even in
the highest temperature case. It is possible to observe some recoherent 
effects in the case c, at the time in which unstable subsystems have
fully decohered. 

When the final system A is an upside-down oscillator \cite{ZP,elze}, an
unstable point forms in the center of the phase space with
associated stable and unstable directions. These are characterized
by Lyapunov coefficients $\Lambda$ ($\Lambda = 2 \omega^2$ in the linear case). The 
time dependence of the package width in the direction of the momenta is given by
$\sigma_p(t) = \sigma_p(t_0) \exp{[\Lambda t]}$, where
$\sigma_p(t_{0})$ is the corresponding width at the initial time. 

Diffusion effects limit the squeezing of the state on the phase space. The
bound on the width of the packets is given by $\sigma_{\rm
c}=\sqrt{2{\cal D}_{\rm i}/\Lambda}$ \cite{JPP-LH,habib} (where
${\rm i}$ is b or d). There is another scale, $t_{\rm max}$
corresponding to the time in which decoherence starts to be
effective, and after which squeezing becomes of the order of the
limiting value. One can estimate the time corresponding to the
transition from reversible to irreversible evolution as
$t_{\rm c} = \frac{1}{\Lambda} \ln{\frac{\sigma_p(0)}{\sigma_{\rm c}}}$. 
Thus, we can use this scale as the 
typical scale for decoherence, setting 
$t_D \approx \frac{1}{\Lambda} \ln{\frac{\sigma_p(t_{\rm max})}
{\sigma_{\rm c}}}$. Therefore we obtain
$t_D =  t_{\rm max} + \frac{1}{\Lambda} \ln{\frac{\sigma_p(0)}{\sigma_{\rm c}}}$.
For the same parameters used in Fig.\ref{figure}, we can  
numerically estimate decoherence times as: $t_{D_{\rm b}} \sim 7.7$ 
and  $t_{D_{\rm d}} \sim 6.4$, for the first set of parameters on 
the left of Fig.\ref{figure} , where $\gamma_0 = 0$; $t_{D_{\rm b}} \sim 2.4$ and
$t_{D_{\rm d}} \sim 2.7$; for $\gamma_0k_B T = 1$, and $t_{D_{\rm
b}} \sim 1.6$ and $t_{D_{\rm d}} \sim 1.7$, in the hight T case
$\gamma_0k_B T = 100$. For the set in Fig. \ref{figure} on the right, 
we obtain: for $\gamma_0 = 0$; $t_{D_{\rm b}}
\sim 3.0$ and $t_{D_{\rm d}} \sim 2.7$. We also got $t_{D_{\rm
b,d}} \sim 0.1$, for $\gamma_0k_B T = 1$, and $t_{D_{\rm b,d}}
\sim 0.6$ in the case $\gamma_0k_B T = 100$. All these results
agree with the decoherence times, defined by the times at which the
decoherence factor $\Gamma (t)$ goes to zero, that can be seen in
the plots above.

Decoherence times for cases (a) and (c) occur as for the usual 
harmonic systems. We can estimate them by using the result of the
high temperature limit of the QBM paradigm,
i.e. $t_D$ is the solution of: $1 \approx L^2 \int_0^{t_D}{\cal
D}(s) ds$ (we have to take the typical distance $L$ as $2\sigma$,
proportional to the dispersion in position of our initial 
packet). The inset in Fig. \ref{figure} on the right, shows $\Gamma (t)$ for a
longer time scale in order to establish the corresponding
hierarchy in the  environmental decoherent effectiveness.

In this article we analyzed the decoherence induced by an
effective environment. The effective environment was considered to
be formed by part of a composite system and an infinite set of
harmonic oscillators. The composite system was considered to be
any of the four possible combinations made up with a harmonic and
an inverted oscillator.

Since a set of harmonic oscillators is a stable system, small
perturbations due to the state of the coupled system do not induce
exploration of a large volume of the phase space for any
oscillator. When one considers an inverted oscillator, it can
explore its volume more efficiently when it is perturbed.

We integrated out subsystem B, in order to study
the effect of having (or not) unstable degrees of freedom into the
full environment. Then we analyzed different situations and
concluded that cases (b) and (d) are the most efficient (smaller
decoherence times) at high temperatures, and (d) is the most
diffusive case, when one turns off the thermal bath. There is a
clear hierarchy between the different compositions of the
composite systems. Those in which oscillator A is unstable (cases
(b) and (d)) decohere before than those with a harmonic oscillator
as the A-subsystem (cases (a) and (c)). At high temperatures of
the external environment, it has been shown that cases (b) and (d)
have the same decoherence time scale, while composite system (c)
losses quantum coherence before case (a).

This work was supported by UBA, CONICET, Fundaci\'on Antorchas, and ANPCyT,
Argentina.

\end{document}